\documentclass[10pt,a4paper,twoside]{article}
\usepackage{epsfig}
\usepackage{baltlat6}
\usepackage{array}
\begin{document}
\ \
\vspace{0.5mm}
\setcounter{page}{1}


\titleb{New observations of the old magnetic nova GQ Muscae}

\begin{authorl}
\authorb{W. Narloch}{1}, 
\authorb{J. Kaluzny}{1}, \authorb{W. Krzeminski}{1},
\authorb{W. Pych}{1}, \authorb{M. Rozyczka}{1},  
\authorb{S. Shectman}{3}, \authorb{I. B. Thompson}{3}, \authorb{T. Tomov}{2}
\end{authorl}

\begin{addressl}
\addressb{1}{Nicolaus Copernicus Astronomical Center,
Bartycka 18, 00-716 Warsaw, Poland;
wnarloch,jka,mnr,wk,pych@camk.edu.pl}

\addressb{2}{Torun Centre for Astronomy, Nicolaus Copernicus University,
 ul. Gagarina 11, 87-100, Torun, Poland; Toma.Tomov@astri.umk.pl}

\addressb{3}{Carnegie Institution of Washington, 813 Santa Barbara Street,
      Pasadena, CA 91101, USA; ian@ociw.edu}
\end{addressl}


\begin{summary} 
We report photometric observations of GQ Mus performed between 1992 and 2011. 
We find that the total amplitude of the orbital modulation of its brightness 
decreased from $\sim$0.9 mag in 1992 to $\sim$0.2 mag in 2010, becoming 
comparable to the amplitude of chaotic flickering on a time scale of several 
minutes. Optical spectra obtained in 2001 and 2012 indicate continuing 
activity of GQ Mus. The spectra show broad emission lines of He~II and H~I 
typical for magnetic cataclysmic variables. The nova was observed as
an UV-bright object in 2001 and 2012.
We also show that the orbital period of GQ Mus has been constant between 
1989 and 2010-2011. 
\end{summary}


\resthead{New observations of the old magnetic nova GQ Muscae}
{W. Narloch, J. Kaluzny, W. Krzeminski, W. Pych, M. Rozyczka, S. Shectman, I. B. Thompson, 
T. Tomov}

\sectionb{1}{INTRODUCTION}

The cataclysmic variable GQ Muscae (Nova Muscae 1983) is a
classical novae. It was discovered by W. Liller on January 18th, 1983 
(Shylaja 1983; Liller \& Overbeek 1983). At the moment of discovery its brightness
was estimated at 7.2 mag. The post-outburst optical light curve 
of GQ~Mus was unusual: after an early decline lasting for about four months  
its brightness remained approximately constant for the next five months
(Krautter et al. 1984). Duerbeck (1987) classified it as a moderately 
fast nova with $t_{3}=45$ d. With a pre-outburst magnitude of $V\geq 21$ 
(Krautter et al. 1984) GQ~Mus belongs to a few known novae whose outburst 
amplitude approaches or even exceeds 14 mag. 

Early spectra of GQ~Mus were characterized by broad and strong $H I$ 
and $He I$ emission lines. Lines of $Fe II$, $Ti II$ and $N III$ 
were also visible, and they indicated a fast expansion of the ejected 
envelope (Charles 1983; Whitelock et al. 1983). Spectra obtained between 
1984 and 1994 showed lines 
of highly ionized atoms, indicating a very high temperature of the central 
source of $2.65\times 10^{5}$K (Diaz et al. 1995; Morisset \& Pequignot 1996). 

Photometric observations obtained in 1988 by Diaz \& Steiner (1989)  
revealed a brightness modulation with a period of 85.5~min. 
The light curve was asymmetric, with the full range reaching 
1 mag in the $V$-band. Further photometric and spectroscopic monitoring
of allowed the classification of GQ~Mus as a magnetic cataclysmic variable 
(Diaz \& Steiner 1990, 1994) and that 
radial velocities measured from strong emission lines of $HeII \lambda4686$, 
$OIV \lambda3411$ and $NIII \lambda5270$ could be phased with the photometric 
period. The natural interpretation is that the 85.5~min periodicity 
corresponds to the orbital period of the system.

GQ Mus was the first classical nova with X-ray emission detected during an 
outburst (\"{O}gelman et al. 1984). It remained in an active 
state for several years longer than predicted -- supersoft X-ray emission 
persisted in 1992 (\"{O}gelman et al. 1993). Later observations 
conducted in 1993 showed that the X-ray flux had decreased by a factor
of 100 within one year (Shanley et al. 1994), indicating the extinction of 
the supersoft X-ray source.

These unusual properties of GQ Mus prompted us to embark on an occasional 
photometric monitoring of this object between 1992 and 2011.
In the present paper we report the results of those observations.
We also present two optical spectra collected in 2001 and 2012. 
The next section summarizes the observations, data reduction 
and calibrations. In section 3 we discuss the main results, and a brief
summary of the paper is given in section 4. 

\sectionb{2}{OBSERVATIONS, DATA REDUCTION AND CALIBRATION}

Photometric observations of GQ Mus were made with 1.0-m Swope and 
2.5-m du Pont telescopes at Las Campanas Observatory in the period 
1992--2011. The TEK2 and SITE3 CCD cameras were used on the Swope telescope, 
while the TEK5 and SITE2K cameras were employed on the du Pont telescope.  
A schematic log of the observations is given in Table 1. The variable was 
observed on 28 nights, yielding a total of 1080 $V$-band frames and 25 
$B$-band frames. In 2001 we also obtained two frames 
in the$U$-band with SITE3/Swope.
 
Profile photometry was extracted using the Daophot/Allstar programs
(Stetson 1987, 1990). The data from SITE3 were additionally reduced with 
the image subtraction code ISIS (Alard \& Lupton 1998; Alard 2000).  
Measurements were transformed to the standard $UBV$ sytem based on observations of 
ten standard stars from two Landolt fields (Landolt 1992) performed on 
June 17/18, 2001. The standard stars were observed at air-masses of 1.15 
and 1.20 while GQ~Mus was observed at an air mass of 1.27. Assuming 
average extinction coefficients for LCO, we obtained for GQ~Mus 
$V=18.65\pm 0.02$ mag, $(B-V)=0.27\pm 0.02 $ mag and $(U-B)=0.78\pm 
0.03$ mag at $HJD=2452078.48$. 
The data from the remaining cameras were transformed to 
the standard $BV$ system based on local standards observed with SITE3.
For most of the differential $V$ photometry the accuracy of 
individual measurements ranges from 0.01 to 0.02~mag. 


\begin{table}[!t]
\begin{center}
\vbox{\footnotesize\tabcolsep=3pt
\parbox[c]{124mm}{\baselineskip=10pt
{\smallbf\ \ Table 1.}{\small\
Summary of photometric observations of GQ Mus.\lstrut}}
\begin{tabular}{ccccccc}
\hline
         &        &         & \multicolumn{2}{c}{No. of frames} &   &     \\ \cline{4-5}
Detector & Season & No. of  & V & B              & Ave. exp. & $<$seeing$>$\\
          &        & nights  &   &                & time in $V$[s] & [arcsec] \\     
\hline
 TEK2     & 1992  & 4 & 165 &    & 185 & 1.5  \hstrut \\
 SITe3    & 1997  & 3 & 21  &    & 122 & 1.3    \\
          & 2000  & 1 & 21  &    & 300 & 3.0   \\
          & 2001  & 3 & 43  & 2  & 300 & 1.5 \\
          & 2002  & 6 & 173 &    & 300 & 1.8   \\
 TEK5     & 2008  & 1 & 5   &    & 60 & 1.5     \\
 SITe2k   & 2010  & 8 & 395 & 19 & 70 & 1.2 \\
          & 2011  & 2 & 257 & 4  & 60 & 1.5
        \lstrut \\
\hline
\end{tabular}
}
\end{center}
\vskip-4mm
\end{table}


A medium resolution spectrum of the nova was obtained with the B\%C 
spectrograph mounted on the Magellan Baade telescope. A 20 min. exposure 
centered on 23h37m UT was made on June 18, 2001. The spectrum 
covered the range 3773-4490 \AA\ at a resolution of 2800, and 
an average signal to noise ratio of~8.
A second medium resolution spectrum of GQ Mus was collected with 
the South African Large Telescope on April 01, 2012. The RS 
spectrograph was used with the PC0900 grating and a 1.5 arcsec slit,  
providing a resolution of $R=800$.
Two 5 min exposures were centered 
on 0h:38m UT. The combined spectrum has a signal to noise ratio of 
2.3 and 2.8 at $H\beta$ and $H \alpha$, respectively. The reduced 
and flux-calibrated spectrum covers the range 3800-6575 \AA\ with 
60-\AA\ wide gaps at 4530~\AA\ and 5590 \AA.

\sectionb{3}{RESULTS}
In Fig. 1 we present a schematic $V$-light curve of GQ Mus based on 
data collected between 1992 and 2011. In that period the average luminosity 
of the nova dropped from $<$V$>$ = 17.87 mag to $<$V$>$ = 19.01 mag. 
Moreover, after 1992 the amplitude of the periodic variability significantly
decreased. 
It is possible that around 2009 the luminosity of the nova stabilized, as no 
systematic change of $<$V$>$ was observed during the subsequent two years. Note 
that in 2010-2011 the nova was still  brighter 
than in the pre-outburst stage by at least two magnitudes.

We have analyzed yearly light curves for periodicity. For three seasons 
the amount of data was sufficient to establish the period with a high accuracy. 
We obtained $P=0.05936538(\pm 8)$~d, $P=0.0593640(\pm 2)$~d and $P=0.05936553(\pm 4)$~d 
for the years 1992, 2002 and 2010-2011, respectively, where the errors are 1-$\sigma$ 
uncertainties of the last significant digit. 
Diaz and Steiner (1994) measured $P=0.0593650(\pm 1)$~d for 1989-1990.
The combined measurements  indicate that  
the orbital period of GQ~Mus was constant between 1989 and 2011 at a level of
$dP/P\leq 1.7\times10^{-6}$.
Phased light curves for 1992, 2002 and 2010-2011 are presented in Fig. 2. The shape of
the 1992 curve is generally similar to that obtained by Diaz and Steiner (1994) for the 
period 1989-1990. However, in 1992 the ascending branch was less steep, and the
bump on the descending branch was less strongly pronounced. The light curves for 2002 
and 2010-2011 have an amplitude of only $\sim$0.1 mag, and are noisy due to flickering 
on a time scale of a few minutes. The light curve for 1994 (Diaz et al. 
1995) showed a significantly reduced amplitude in comparison with those for the
1989-1990 
and 1992 seasons. Apparently the orbital modulation weakened with time, 
 becoming comparable to the amplitude of flickering by the 
 observing 2010-2011 seasons.

With galactic coordinates $l=297.2$ deg and $b=-5.0$ deg, the nova resides in a rather 
crowded stellar field. For a heliocentric distance of $3.2-4.8$~kpc 
(Diaz \& Steiner 1994) it 
is located 280-420~pc above the galactic disk, so that it must be either a thick disk 
or a halo object. Figure 3 shows $V/(B-V)$ and $V/(U-B)$ color-magnitude diagrams (CMD)
of the $3\times 5$~arcmin$^2$ field containing GQ~Mus. The nova 
is the bluest object on both CMDs, and the only UV-bright star in the analyzed field.
While the $V$-magnitude increased by about 0.4 mag between
2001 and 2010, the object became  redder by just 0.06~mag.   

Figure 4 shows the spectrum taken in June 2001, smoothed with a 7-pixel moving 
box. While smoothing reduces the resolution, it makes the broad Balmer emission 
lines more clearly visible. 
Such broad Balmer lines have been  observed in several polars and intermediate polars 
in an active state (Rosen et al. 1993; Piirola et al. 2008; Thomas et al. 2012).
The spectrum obtained with the SALT telescope on April 2012, smoothed with a 9-pixel 
moving box, is presented in Fig. 5. It was flux calibrated and then normalized
at 5500 \AA.
Balmer lines together with $HeII\lambda4686$ and $NIII\lambda4650$ are seen in emission, with 
$H\alpha$ being particularly strong. 
The spectrum still exhibited signatures typical for magnetic cataclysmic 
variables in 2012. Moreover, the spectrum implies a very blue color for the nova 
(no correction for interstellar  reddening has been made, so that GQ Mus was  
even brighter in the UV than the slope of the 
spectrum suggests). 


\sectionb{4}{CONCLUSIONS}
Almost three decades after the outburst of GQ Mus the object was 
brighter than in the pre-outburst stage by at least two magnitudes. Its light 
curve exibited  orbital modulation with a shape typical 
for magnetic CVs. In both  2001 and 2012 the nova was bright in the UV. 
Spectra obtained in 2001 and 2012 show broad emission lines of H and He, 
characteristic for polars.
We found no evidence for period changes between 1989 and 2011 at an upper 
limit of $dP/P\leq 1.7\times10^{-6}$.

\thanks{This project was supported by grant DEC-2012/05/B/ST9/03931 
from the Polish National Science Center.}

\References

\refb Alard C., Lupton R. H., 1998, ApJ, Vol. 503, p. 325

\refb Alard C., 2000, A\&A, 

\refb Charles P., 1983, IAU Circ., No. 3766


\refb Diaz M. P., Steiner J. E., 1989, ApJ, 339, 41-43

\refb Diaz M. P., Steiner J. E., 1990, Rev. Mex. Astron. Astrof., 21, 369-372

\refb Diaz M. P., Steiner J. E., 1994, ApJ, 425, 252

\refb Diaz M. P., Williams R. E., Phillips M. M., Hamuy M., 1995, MNRAS, 277, 959-964

\refb Duerbeck H. W., 1987, SSRV, 45, 1-212



\refb Krautter, J. et al., 1984, A\&A, 137, 307


\refb Landolt A. U., 1992, AJ, 104, nr 1, 340

\refb Liller W., Overbeek M. D., 1983, IAU Circ., No. 3764

\refb Morisset C., Pequignot D., 1996, A\&A, 312, 135-159

\refb \"{O}gelman H., Beuermann K., Krautter J., 1984, ApJ, 287, 31-34

\refb \"{O}gelman H., Orio M., Krautter J., Starrfield S., 1993, Nature 361, 331

\refb Piirola V., Vornanen T., Berdyugin A., Coyne S. J. G. V., 2008, ApJ, 684, 558

\refb Rosen S. R., Mittaz J. P. D., Hakala P. J., 1993, MNRAS, 264, 171

\refb Shanley L., Gallagher J. S., \"{O}gelman H., Orio M., Krautter J., 1994, AAS, 184, 4601

\refb Shylaja B. S., 1983, The Observatory, Vol. 103, p. 203-204

\refb Stetson P. B., 1987, PASP, 99, 191-222

\refb Stetson P. B., 1990, PASP, 102, 932-948

\refb Thomas H. C., Beuermann K., Reinsch K., Schwope A. D., Burwitz, V., 2012, A\&A, 546, 104

\refb Whitelock P. A., Carter B. S., Feast M. W., Glass I. S., Laney D., 
Menzies J. W., Walsh J., Williams P. M., 1984, MNRAS, 211, 421-432


\begin{figure}[!H]
\vbox{
\centerline{\psfig{figure=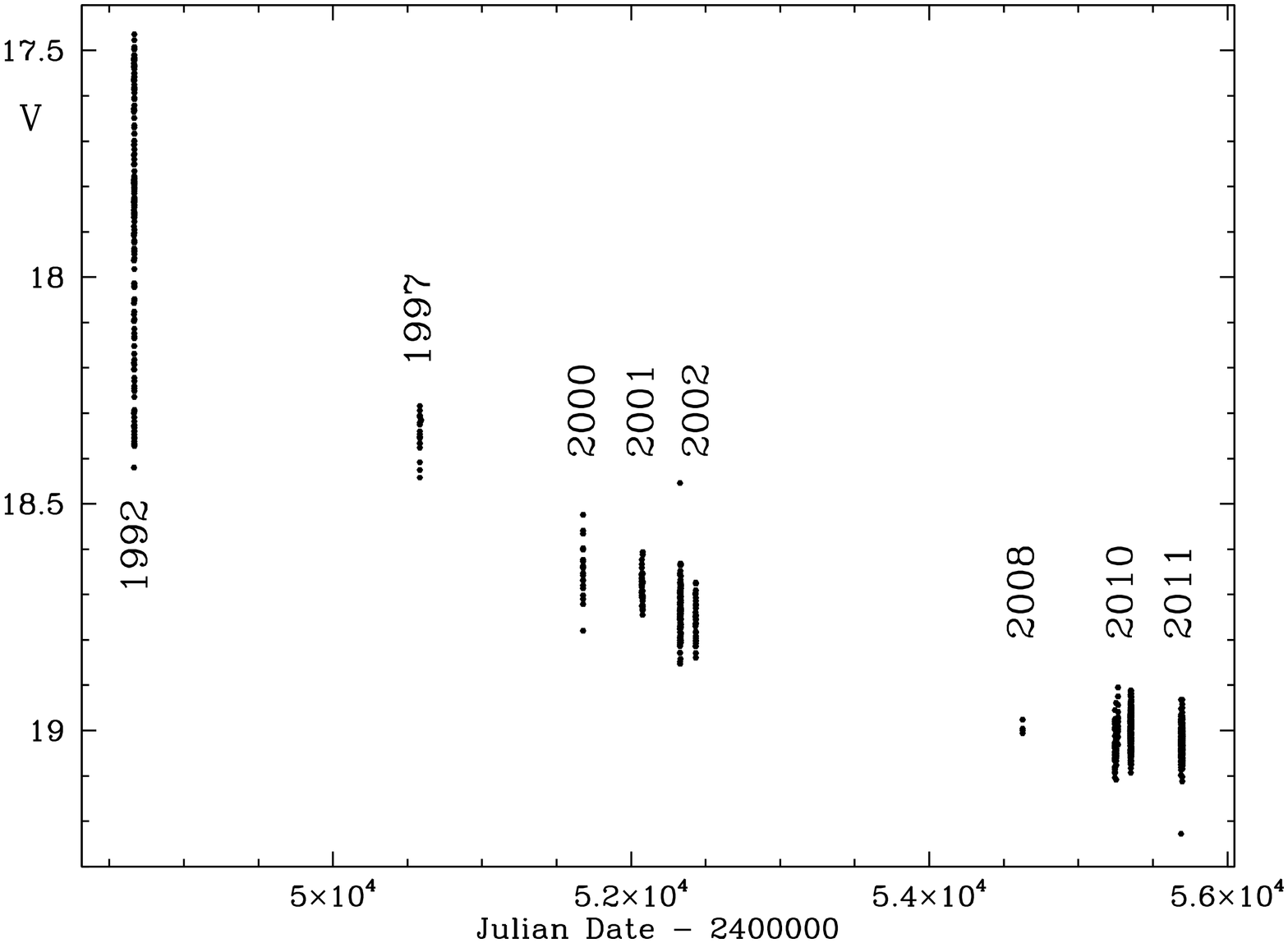,width=100mm,angle=270,clip=}}
\vspace{1mm}
\captionb{1}
{The Light curve of GQ Mus for the period 1992-2011.}
}
\end{figure}


\begin{figure}[!tH]
\vbox{
\centerline{\psfig{figure=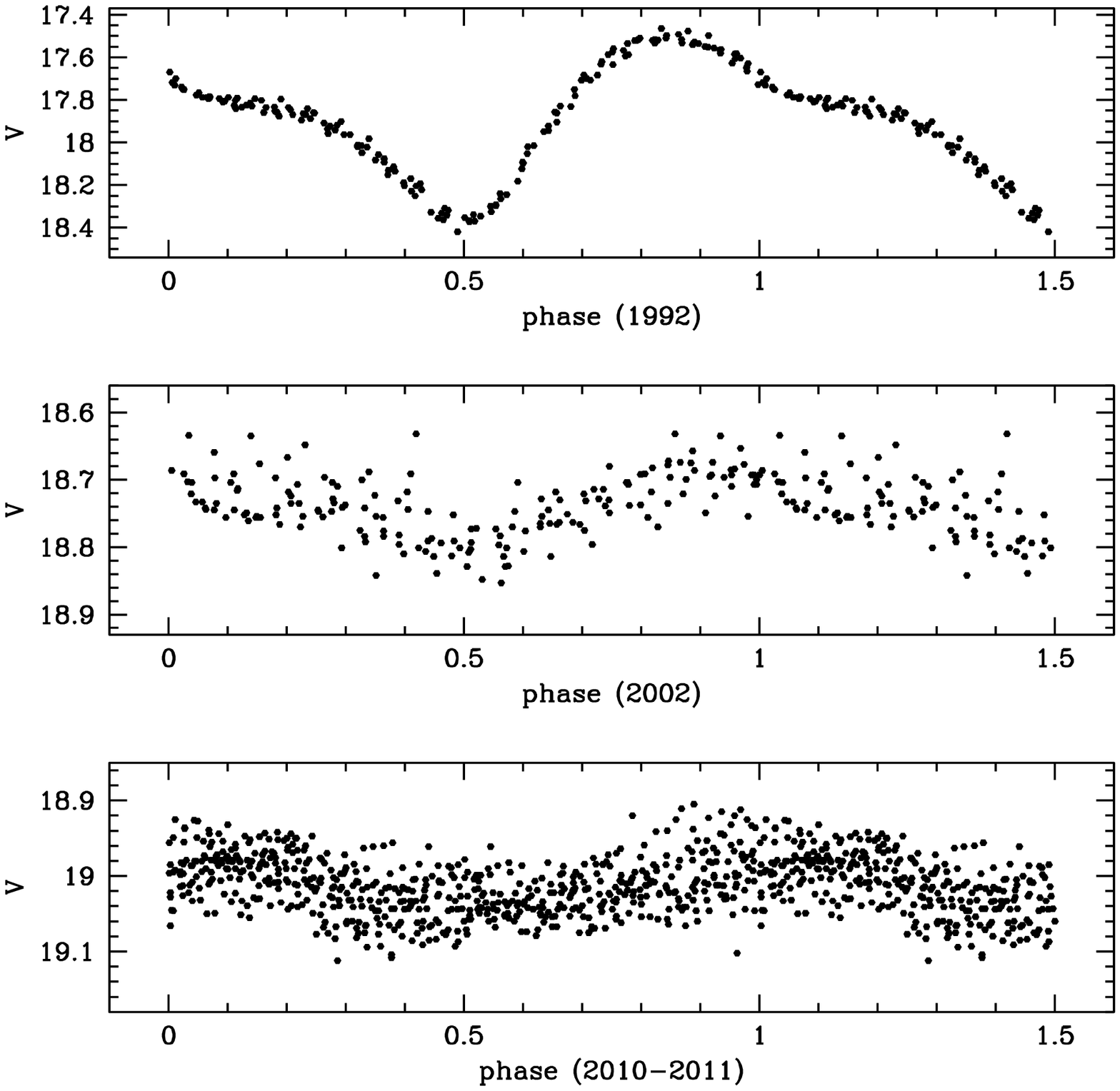,width=100mm,angle=0,clip=}}
\vspace{1mm}
\captionb{2}
{Phased light curves of GQ Mus in $V$ for seasons 1992, 2002 and 2010-2011.}
}
\end{figure}


\begin{figure}[!tH]
\vbox{
\centerline{\psfig{figure=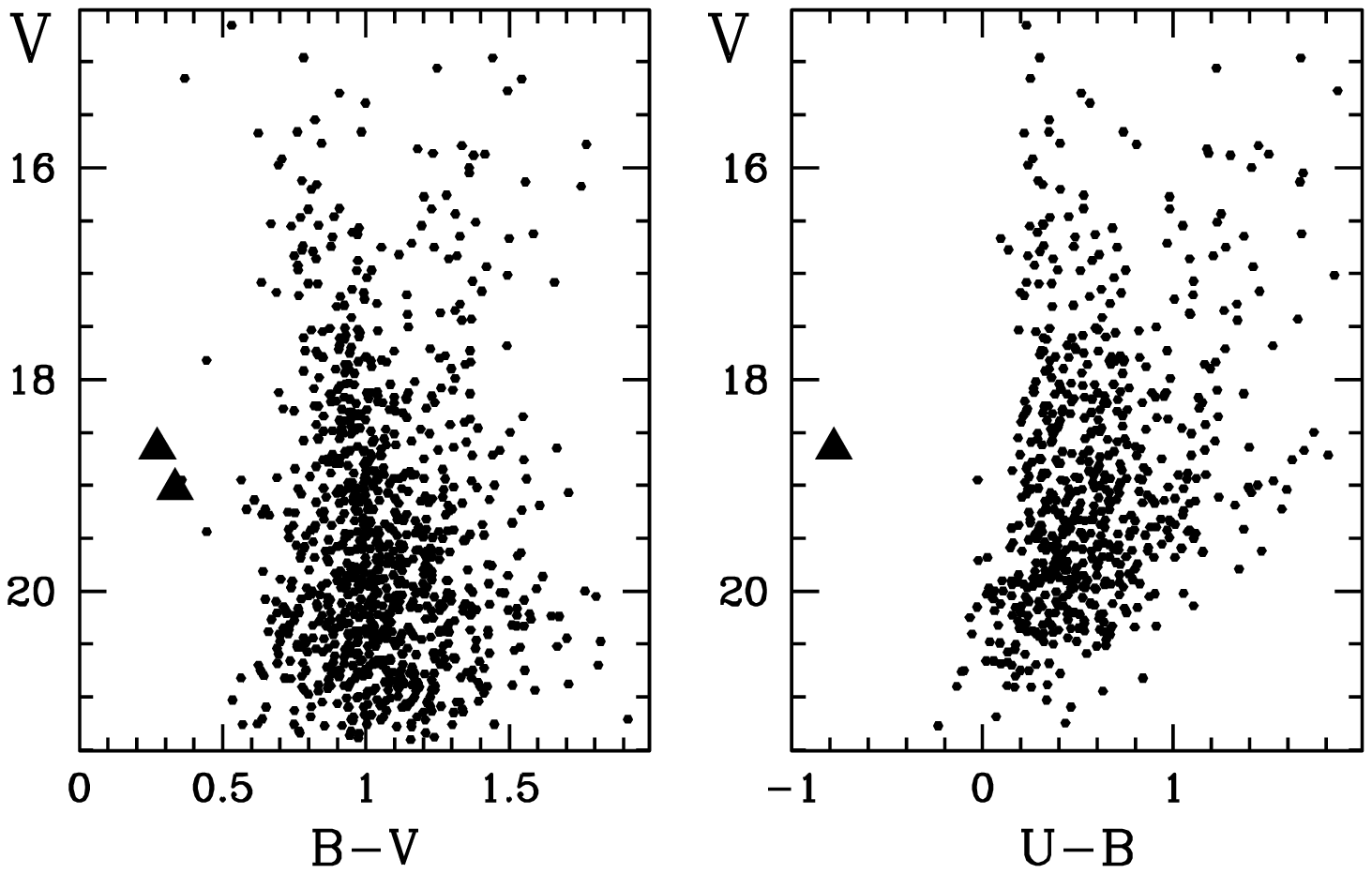,width=150mm,angle=0,clip=}}
\vspace{1mm}
\captionb{3}
{Color-magnitude diagrams for {\bf the 3$\times$5 arcmin$^2$ field centered on GQ Mus}.
Positions of the variable are marked with triangles, with $V$ = $\sim$18.5 mag corresponding 
to the season 2001, and $V$ = $\sim$19.0 mag to the season 2010.}
}
\end{figure}




\begin{figure}[!tH]
\vbox{
\centerline{\psfig{figure=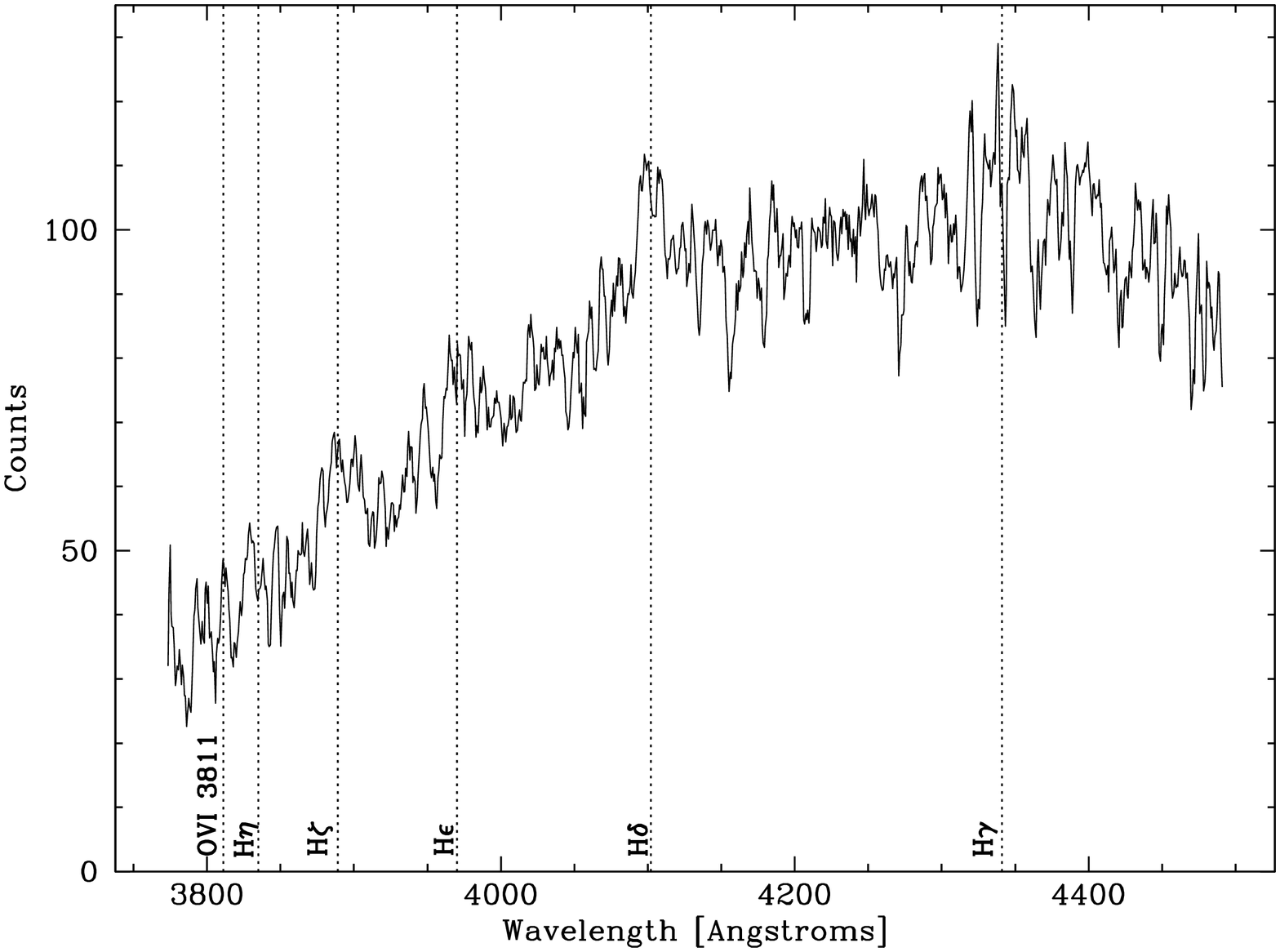,width=100mm,angle=270,clip=}}
\vspace{1mm}
\captionb{4}
{The spectrum of GQ Mus from June 18, 2001 smoothed with a 7-pixel moving box.
}
}
\end{figure}

\begin{figure}[!tH]
\vbox{
\centerline{\psfig{figure=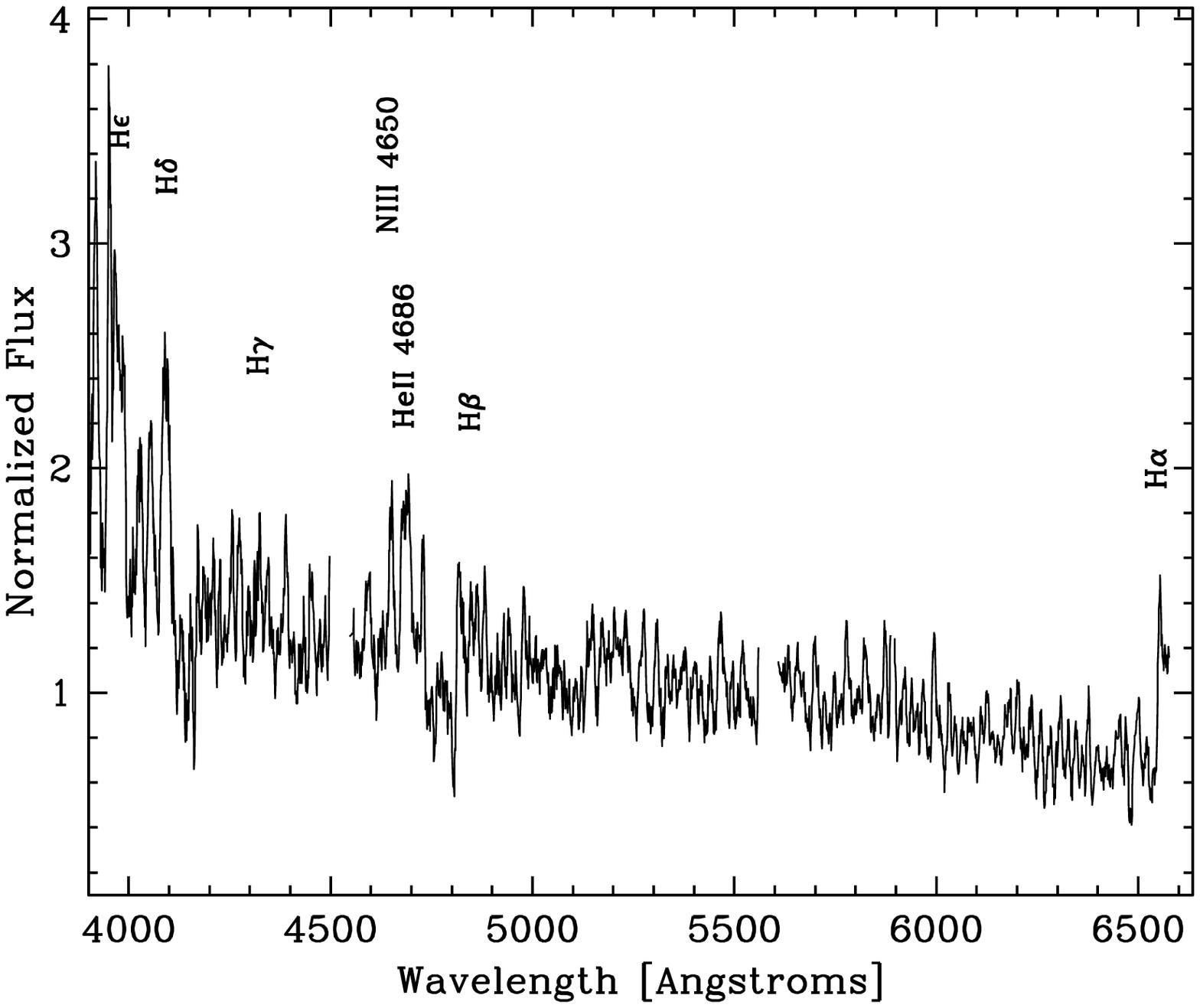,width=120mm,angle=0,clip=}}
\vspace{1mm}
\captionb{5}
{The spectrum of GQ Mus from April 2012 smoothed with a 9-pixel moving 
box.}
}
\end{figure}


\end{document}